# A Simulation Approach for Determining the Spectrum of DNA Damage Induced by Protons

**Mojtaba Mokari [1, 2], Mohammad Hassan Alamatsaz [1], Hossein Moeini [1], and Reza Taleei [3]**

*1) Department of Physics, Isfahan University of Technology, Isfahan 84156-83111, Iran*
*2) Faculty of Science, Behbahan Khatam Alanbia University of Technology, Behbahan 63616-47189, Iran*
*3) Division of Medical Physics and Engineering, Department of Radiation Oncology, UT Southwestern Medical Center, Dallas, TX 75390-8542, USA*

Electronic mail: mokari@bkatu.ac.ir

## Abstract

In order to study the molecular damage induced in the form of single-strand and double-strand breaks by the ionizing radiation at the DNA level, Geant4-DNA Monte Carlo simulation code for complete transportation of primary protons and other secondary particles in liquid water have been employed in this work. To this aim, a B-DNA model and a thorough classification of DNA damage concerning their complexity were used. Strand breaks were assumed to have been primarily originated by direct physical interactions via energy depositions, assuming a threshold energy of 17.5 eV, or indirect chemical reactions of hydroxyl radicals, assuming a probability of 0.13. The simulation results on the complexity and frequency of various damage are computed for proton energies of 0.5 to 20 MeV. The yield results for a cell (Gy.cell)$^{-1}$ are presented, assuming 22 chromosomes per DNA and a mean number of 245 Mbp per chromosome. The results show that for proton energies below 2 MeV, more than 50% of energy depositions within the DNA volume resulted in strand breaks. For double-strand breaks (DSBs), there is considerable sensitivity of DSB frequency to the proton energy. A comparison of DSB frequencies predicted by different simulations and experiments is presented as a function of proton LET. It is shown that, generally, our yield results (Gy.Gbp)$^{-1}$ are comparable with various experimental data and there seems to be a better agreement between our results and a number of experimental studies, as compared to some other simulations.

### Keywords

## Introduction

When ionizing radiations interact with cells, they will cause both early and late biophysical effects. Early effects could last for a few femtoseconds to few days, while the late effects would last for years. Initial effects of radiations include the effects from physical processes due to the ionization and excitation interactions as well as the effects of the chemical radicals [1]. The interaction of the ionizing radiations with cells that would instigate DNA damage has major role in cancer therapy. DNA damages include single-strand and double-strand breaks (SSB and DSB). Double-strand breaks can cause the death of the cell, if they result in a mis-repair or unrepair [2].

 Until now an accurate quantitative study of various types of damage e.g. complex type DSB has not been possible [3]. As such, people try to use simulation methods to study the biophysical interactions of radiations. Most common Monte Carlo simulation codes use the spatial history of the particles inside the matter that are implemented in codes such as FLUKA [4], Geant4 [5] [6], MCNP [7], MCEP [8], PITS [9], PENELOPE [10], or Monte Carlo Track Structure (MCTS) codes like PARTRAC [11] and KURBUC [12] [2]. In calculating the damage, parameters such as the type of the incident radiation, radiation energy, interaction cross sections, threshold energy $E_{ssb}$ [13], and the probability of indirect interactions of chemical radicals with the DNA [14] appear to be important. In the MCDS code, which has a quasi-phenomenological algorithm, one avoids the initial simulation of the physical and chemical processes [15]. To calculate damage frequency using codes such as KURBUC or Geant4-DNA, one first simulates the ionizing radiations in the environment and subsequently calculates the damage considering the energy deposits and the probabilistic effect of the chemical radicals. KURBUC is a comprehensive code for simulating various types of particles such as electron, proton, neutron, and heavy ions. KURBUC-Proton was the first MCTS code and released in 2001 for simulating the proton interactions, which covers the proton energy ranges of 1 keV to 1 MeV with the unique feature of including charge-exchange model for $H^+$, $H^0$ and $H^-$ in the Bragg peak region [16]. Geant4-DNA is a subproject of Geant4 – a general purpose particle-matter Monte Carlo simulation toolkit – that is extended with processes for modeling the early biological damage induced by ionizing radiation at the DNA scale. The implemented models and physics processes in Geant4-DNA allow for step-by-step simulation of the interaction of particles in liquid water down to the eV energy scale [17]. Meylan *et al* [18] simulates fibroblast cell nucleus using Geant4-DNA with protons. Lampe *et al* [19] effectively simulates the bacterial nucleus and studies the DNA damage from electrons and protons in a modelled full genome of an Escherichia coli cell using Geant4-DNA.

In this work, we have used Geant4-DNA (Geant4 version 10.3) to simulate primary protons, with energies ranging from 0.5 MeV to 20 MeV, and secondary particles in liquid water. Herewith, we compared our calculations to the published experimental data. The objective of this article is to calculate the initial direct and indirect DNA damage by incident protons and secondary particles, taking into account both physical and chemical interactions. For this purpose, the frequency of simple and clustered damages of different complexity as well as the yield for SSB and DSB are determined. Yield values are calculated per nucleotide pair (base pair) and per cell and the results are compared with other simulation works and experimental studies of Frankenberg *et al* [20], Belli *et al* [21], Belli *et al* [22], and Campa *et al* [23]. Calculation of the yield per nucleotide pair is done in two ways. One is based on the total number of SSBs and DSBs, whereas in the other method we used the frequency of the deposited energies.

## Methods

When a cell is irradiated, the initial damage in the DNA molecules would be the result of direct physical interactions of the ionizing radiations with the DNA or indirect chemical reactions of the produced free radicals in the surrounding water. The simulation study in this work has been done using the Geant4-DNA toolkit for particle transport through water. Taking into account both physical and chemical interactions, the toolkit allows extracting information such as energy deposition, position and time of flight of various particles and radiations at different steps of their propagation through matter. On their path through matter, the primary protons and secondary particles can undergo various processes. The cross sections used in Geant4-DNA account for elastic scattering, ionization, excitation, and Auger cascades, which are of importance in our DNA damage simulations. The energy cut-off threshold for tracking electrons is 7.4 eV, by the default of Geant4-DNA, below which the tracking would be ceased and the remaining kinetic energy of the electron would be deposited at once.

Using a PDB file, we extracted the position of the atoms of a 216 bp long double helix B-DNA (equivalent to 73.44 nm and consisting of 432 nucleotides). A B-DNA molecule is one of the common DNA structures in living creatures [24] [25]. Each nucleotide comprises sugar-phosphate groups and a base group (see figure 1 for an artist view of such DNA shapes). We then sampled the DNA molecules within the volume of a water sphere of 100 nanometers radius. Our isotropic point source was located at the center of the water sphere that emitted protons of desired energy. The sampling of DNAs within the spherical volume was done based on the μ-randomness method [26]. We found an optimized number of DNAs which could satisfy the two sampling accuracy criteria explained by Nikjoo *et al* [27] [28]. For the first criterion, we compared the ratio of the energy deposition in the original sphere to its volume with the ratio of energy deposition in the DNAs to their volumes. For the second criterion, the mean of the inverse frequency-averaged mean specific energy $(\bar{Z}_f)^{-1}$ was calculated and compared with the frequency of hits of any size ($f$ ( > 0)) [28]. For both tests, in order to conduct a good sampling, the ratios were targeted to be equal within 5% uncertainty.

The calculations for this study comprised of three stages. The physical stage was the first stage, where the simulation of physical interactions of various particles in water was pursued until they reached the energy or geometrical cut-off. In the second stage, hereafter referred to as the chemical stage, the simulation of physico-chemical and chemical processes up to 1 ns was performed. In the third stage, referred to as the damage formation stage, a written algorithm determined the damage types according to the definition of damage spectra given by Nikjoo *et al* [13].

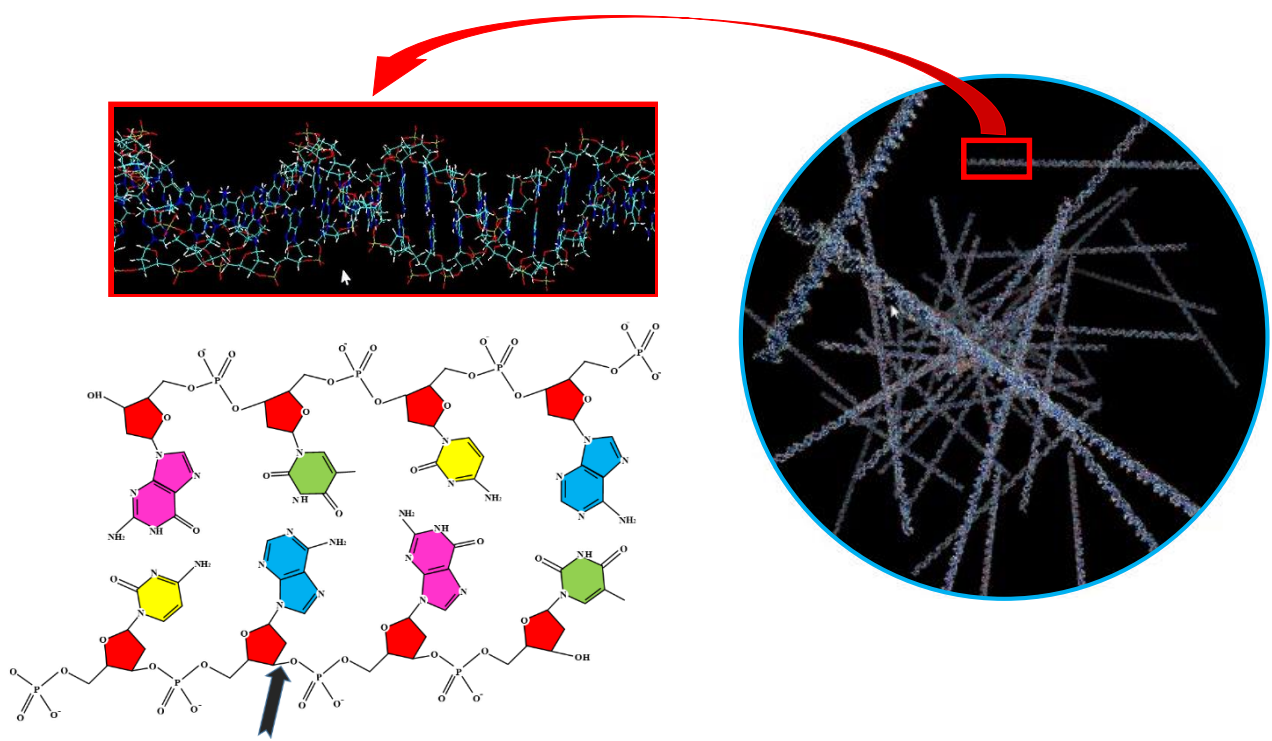

**Figure 1.** 50 DNA segments randomly distributed within the spherical water environment (right) drawn using VMD together with a zoomed-in view of part of a single DNA segment (top left). On the bottom left, a schematic view of part of the DNA molecule containing base, sugar, and phosphate chains is shown. The arrow indicates a representative position at which a single C-O chain break could happen due to the nearby energy depositions [29].

At the end of the physical stage, when the tracking of various particles is completed, the positions and deposited energies, at the end of each step involving ionization or excitation, were derived. For the sum of such energy depositions in one sugar-phosphate volume that was more than a threshold value $E_{ssb}$, a strand break (SB) was registered at the corresponding DNA segment. In accordance with the experimental studies by Martin and Haseltine [30], Terrissol [31], and Kandaiya *et al* [32], the threshold value was assumed to be $E_{ssb} = 17.5$ eV.

For the chemical stage, we have exploited the *TimeStepAction* class of the Geant4-DNA, where we recorded down the position of the produced radicals in the environment 1 ns after throwing the primary proton (hereafter referred to as the chemical stage simulation time). Among the various radicals and molecules that are produced within the water environment, including $H_2O_2$, $H_2$, $e_{aq}^-$, $OH^-$, $OH^\bullet$, $H^+$ and $H^\bullet$, hydroxyl radicals ($OH^\bullet$) would occur more commonly due to their capability to interact with the DNA segments [33] [34]. Table 1 makes a comparison between the simulation and experimental outcomes of the reaction rates for a few chemical reactions involving the production of different radicals. It can be seen that overall, in Geant4-DNA more hydroxyl radicals would react in the environment and hence the share of indirect damages is assumed higher as compared to experiment. The derived positions of the hydroxyl radicals were then checked in our algorithm to see whether they would fall within the volume of any imaginary cylinder of (8 + 2.3) nm diameter, with its longitudinal axis coinciding with the axis of the DNA cylinder of 2.3 nm diameter. For the corresponding DNAs of those cylinders that can pass this condition, the closest sugar or phosphate nucleotide to the hydroxyl radical would then be found and a SB at that DNA segment would be registered with a probability of $P_{OH} = 0.13$ [35] [13]. Following the approach of Nikjoo *et al* [13], we increased the number of events from $10^3$ to $5\times10^3$.

**Table 1.** Chemical interactions and reaction rates in the production of radicals according to Geant4-DNA [36] and experiment [37]. For a more comprehensive list of reactions, check [36] [37].

| Reaction | Reaction Rate (Geant4-DNA) $(dm^3mol^{-1}s^{-1})$ | Reaction Rate (Experiment) $(dm^3mol^{-1}s^{-1})$ |
|---|---|---|
| $H_2 + OH^{\bullet} \rightarrow H^{\bullet} + H_2O$ | $4.17 \times 10^7$ | $4.5 \times 10^7$ |
| $OH^{\bullet} + OH^{\bullet} \rightarrow H_2O_2$ | $0.44 \times 10^{10}$ | $0.6 \times 10^{10}$ |
| $H_2O_2 + e_{aq}^{-} \rightarrow OH^{-} + OH^{\bullet}$ | $1.41 \times 10^{10}$ | $1.3 \times 10^{10}$ |
| $H^{\bullet} + OH^{\bullet} \rightarrow H_2O$ | $1.44 \times 10^{10}$ | $2.0 \times 10^{10}$ |
| $OH^{\bullet} + e_{aq}^{-} \rightarrow OH^{-}$ | $2.95 \times 10^{10}$ | $2.5 \times 10^{10}$ |

As stated earlier, each physical and chemical interaction (in this work, corresponding to satisfying the $E_{ssb}$ and $P_{OH}$ conditions) is considered to cause one SB. The classification of the DNA strand breaks is done according to either the complexity of the clustered damage or the origin of the breaks (direct and indirect). The categorization of the DNA damage types is done according to Nikjoo *et al* [2]and shown in figure 2, using an algorithm written in Python programming.

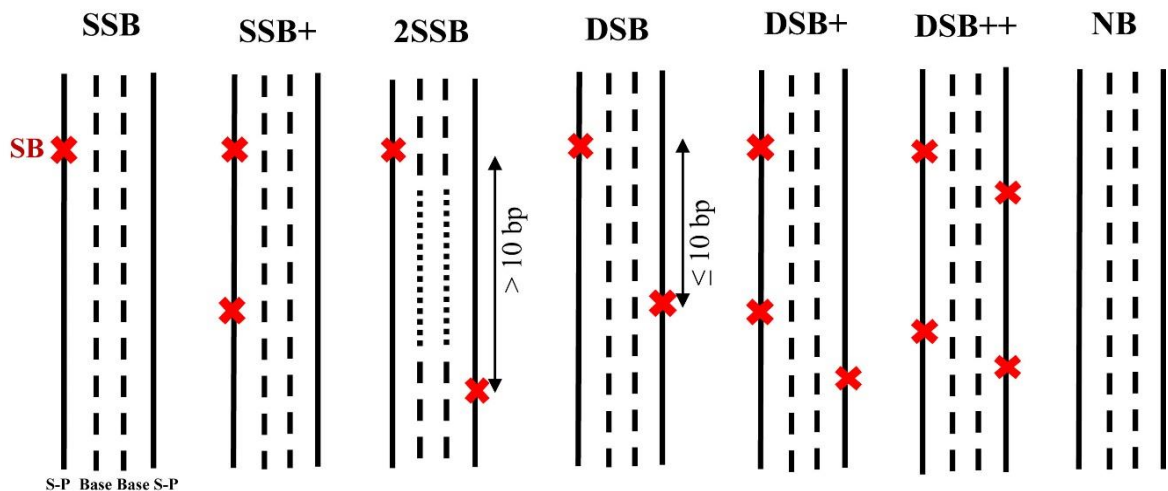

**Figure 2.** Types of the DNA damage induced by direct energy deposition and reaction of OH radicals. For simplicity, the DNA is shown as four parallel lines. The solid lines represent the sugar-phosphate (S-P) backbone and the dashed lines represent the bases. A '×' represents an SB in DNA. If only one '×' is on either of the strands, it will be labeled as SSB. If two '×' are on opposite strands within 10 bp of each other, it will be labeled as DSB. If two SSBs (one on each strand) are more than 10 bp apart, it is labeled as 2SSB, while if two SSBs are within 10 bp apart, but on the same strand of DNA, it is labeled as SSB+. A double-strand break accompanied by additional single strand break(s) within 10 bp separation is labeled as DSB+. More than one double-strand break in the DNA is labeled as DSB++. The NB (no break) category refers to a DNA without any SB [38] [2].

**Results**

Table 2 shows the calculated relative yields of different types of strand breaks as a function of proton energies. In this table, complex damage has been defined as $SSB_c$ (= $SSB^+$ + 2SSB) and $DSB_c$ (= $DSB^+$ + $DSB^{++}$) [39]. These data show that at low energies the majority of proton hits in DNA do lead to damage in the form of strand breaks. Specifically, this is about more than 50% of hits, for proton energies less than 2 MeV. It also shows that the frequency of simple SSB is independent of primary proton energy, comprising about one third of hits. It is interesting to note that 2SSB appears to be the second most frequent type of damage and has a frequency of more than three times that of the $SSB^+$ for all proton energies. This is contrary to the damage frequency from primary electrons Nikjoo *et al* [13], where the frequency of 2SSB is substantially less than $SSB^+$ for various energies. It is also observed that the yield of $DSB^+$ and $DSB^{++}$, are comparable to simple DSB at proton energies lower than 2 MeV. Furthermore, the yields of DSB are about two times lower than those of 2SSB. This is especially interesting considering the definitions of 2SSB and DSB damage, based on the 10 bp distance between the two breaks. It can be seen that, for high proton energies above 2 MeV, they decrease with energy relatively close to each other; whereas for lower primary energies, $SSB_c$ appears to be about 10% less than $DSB_c$. The data in this Table do not account for the role of base damage in inducing DNA strand breaks or adding to their complexities. Hence, these data can only represent a lower limit of the complexity of damages.

**Table 2.** Relative yield of strand breaks classified by complexity as a function of proton energy.

| Energy MeV | $LET_\infty$ keV/μm | NB % | SSB % | SSB+ % | 2SSB % | DSB % | DSB+ % | DSB++ % | $SSB_c$ % | $DSB_c$ % | $Y_{SSB}$ (Gy.Gbp)⁻¹ | $Y_{DSB}$ (Gy.Gbp)⁻¹ |
|---|---|---|---|---|---|---|---|---|---|---|---|---|
| **0.5** | 39.7 | 26.59 | 29.26 | 4.06 | 17.42 | 10.27 | 7.70 | 4.68 | 42.33 | 54.63 | 39.05 | 7.80 |
| **1** | 24.2 | 37.92 | 31.16 | 3.52 | 13.75 | 7.59 | 4.90 | 1.18 | 35.66 | 44.45 | 50.92 | 7.77 |
| **2** | 13.9 | 50.44 | 30.34 | 2.59 | 9.93 | 4.67 | 1.76 | 0.28 | 29.19 | 30.42 | 62.74 | 6.36 |
| **10** | 3.4 | 63.29 | 27.35 | 1.57 | 5.19 | 2.10 | 0.47 | 0.03 | 19.81 | 19.40 | 73.45 | 4.3 |
| **20** | 1.9 | 67.92 | 25.27 | 1.10 | 3.95 | 1.50 | 0.25 | 0.01 | 16.67 | 14.65 | 75.15 | 3.50 |

Figure 3 compares our results for DSB yield with other simulation and experimental data, as a function of the proton LET in water. Though there is considerable difference between our results and the simulation works of Nikjoo *et al* [39] and Friedland *et al* [40], the estimated yields (for LET values below about 30 keV/μm) are reasonably close to the ones estimated by the simulations of Meylan *et al* [18]. Also, the trend of the damage yields for various simulations and experimental data are similar. Our results show a better agreement to Belli et al. [21] and Campa *et al* [23] experimental results, and reproduce the data of Frankenberg *et al* [20] at LETs close to 25 keV/μm.

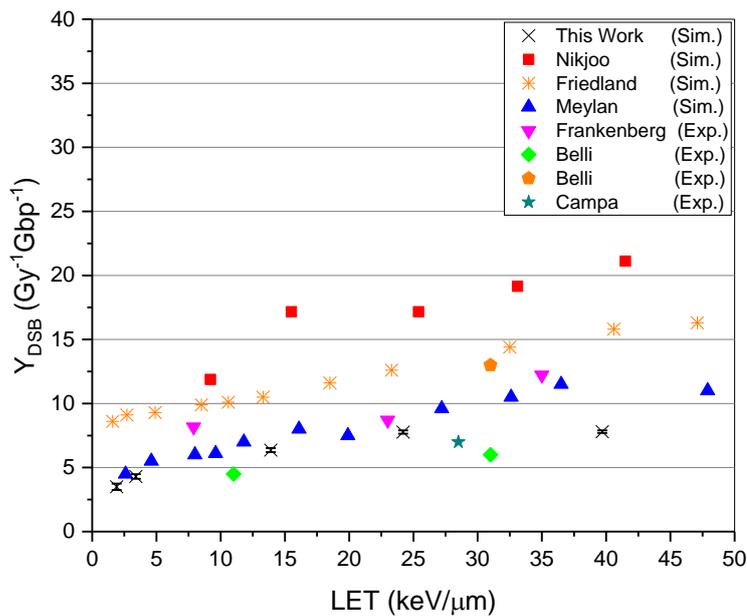

**Figure 3.** Comparison of our simulation results for $Y_{DSB}$ (×) with simulation and experimental results of Nikjoo *et al* [39], Friedland *et al* [40], Meylan *et al* [18], Frankenberg *et al* [20], Belli *et al* [21] (diamonds), Belli *et al* [22] (pentagon), and Campa *et al* [23].

Figure 4 makes a comparison of the frequency of the direct and indirect damages (as the percentage of the total number of breaks), as a function of number of damage sites. For example, for 0.5 MeV protons, the probability for direct interactions capable of inducing one break per DNA segment is about 31.2% of the total breaks. In the case of 0.5 MeV protons with a single damage site, the obtained value for the frequency of All Breaks appears to be less than the corresponding values for Direct and Indirect Breaks. This can be explained as follows. The frequency of All Breaks with single damage site is the sum of the frequencies of Direct and Indirect Breaks with single damage sites minus two times the frequency of damages that are labeled as both single Direct and single Indirect Breaks. Hence for 0.5 MeV protons, assuming that the frequencies of All, Direct, and Indirect Breaks with single damage site are respectively about 29.7%, 31.2%, and 32.4%, the frequency of damage sites that are tagged as both single Direct and single Indirect Break is obtained to be about 16.9%. Thus, for low enough proton energies, say 0.5 MeV, it is more probable to create both single Direct and single Indirect Breaks (16.9%) than either single Direct (31.2%-16.9%=14.3%) or single Indirect Break (32.4%-16.9%=15.5%). As a comparison with other available data, the results for total number of breaks and 1 MeV protons agree closely with the ones in Nikjoo *et al* [39]. As a general feature of these diagrams, the probability of creating direct or indirect single-site damage (All Breaks) significantly increases with increasing the proton energy. For the probability of creating direct or indirect multiple-site damages, there is however not much sensitivity observed to the proton energy. The probability of creating both single Direct and single Indirect Breaks for other proton energies 1, 2, and 10 MeV are obtained from figure 4 to be about 18.4%, 17.9%, and 14.9%, respectively.

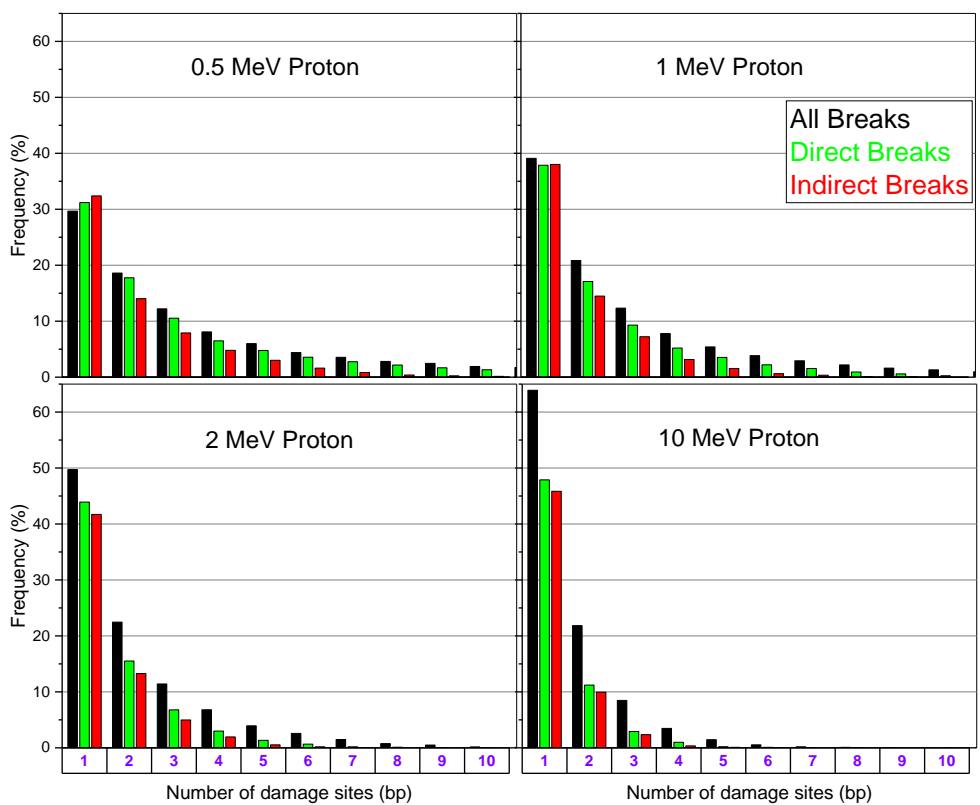

**Figure 4.** Comparison between the relative frequency distributions of the number of damage sites induced by direct and indirect damages, as the percentage of the total number of breaks.

Table 3 lists the number of different types of breaks as function of the deposited energy in the DNA segment. The two columns on the left represent the energy deposition intervals and the corresponding number of events falling within each interval (total hits). The body of the Table indicates the number of events within an energy interval, which produced a specific type of break in our simulations. Thus, of the total 1913 events which deposited between 100 and 150 eV, 739 (39%) produced a SSB. The total number of SSBs created by energy depositions in this energy interval is calculated to be $739 + 168 + 2 \times (563 + 261 + 52 + 2) = 2663$, in which each DSB amounts to two SSB ($SSB_{all} = SSB + SSB^+ + 2 \times (2SSB + DSB + DSB^+ + DSB^{++})$ and $DSB_{all} = DSB + DSB^+ + DSB^{++}$ [41]. Hence, the deposited energy in this interval creates 1.39 $SSB_{all}$ (= 2663 / 1913) per event of 216 bp size. Similarly, for DSBs, a total of 0.16 $DSB_{all}$ (= 315 / 1913) per event are produced in this range of energy deposition.

**Table 3.** Frequency of hits and various types of breaks as function of the deposited energy in DNA segments of 216 nucleotides length, for 2 MeV primary protons.

| Deposited Energy (eV) | Total hits | NB | SSB | SSB+ | 2SSB | DSB | DSB+ | DSB++ |
|---|---|---|---|---|---|---|---|---|
| 0 - 20 | 9933 | 8649 | 1211 | 9 | 56 | 8 | 0 | 0 |
| 20 - 40 | 4627 | 2867 | 1569 | 26 | 144 | 21 | 0 | 0 |
| 40 - 60 | 2744 | 979 | 1471 | 37 | 203 | 53 | 1 | 0 |
| 60 - 80 | 1911 | 430 | 1075 | 64 | 275 | 61 | 6 | 0 |
| 80 - 100 | 1351 | 173 | 725 | 72 | 291 | 84 | 6 | 0 |
| 100 - 150 | 1913 | 128 | 739 | 168 | 563 | 261 | 52 | 2 |
| 150 - 200 | 1076 | 19 | 266 | 105 | 383 | 215 | 81 | 7 |
| 200 - 250 | 636 | 2 | 78 | 77 | 225 | 172 | 76 | 6 |
| 250 - 300 | 369 | 0 | 21 | 28 | 124 | 111 | 70 | 15 |
| 300 - 350 | 198 | 0 | 4 | 14 | 43 | 68 | 60 | 9 |
| 350 - 400 | 80 | 0 | 0 | 4 | 12 | 23 | 29 | 12 |
| 400 - 450 | 35 | 0 | 0 | 1 | 5 | 8 | 15 | 6 |
| > 450 | 128 | 0 | 37 | 8 | 30 | 22 | 22 | 9 |

Figure 5 represents the results of Table 3 and for other proton energies than 2 MeV. The damage yield is calculated for 1000 events and are normalized to the number of hits. One general feature in these graphs is that as the complexity of the breaks increases, the frequency patterns extend and peak at higher deposited energy. Also, for each break type, the lower the proton energy, the higher the amount of the extension and the number of breaks appear to be (corresponding to the integral of the frequency graph). Table 4 shows the frequency distribution of the total number of hits (see Table 3, left column) per unit dose per DNA segment, as a function of the total energy deposited in the DNA segment.

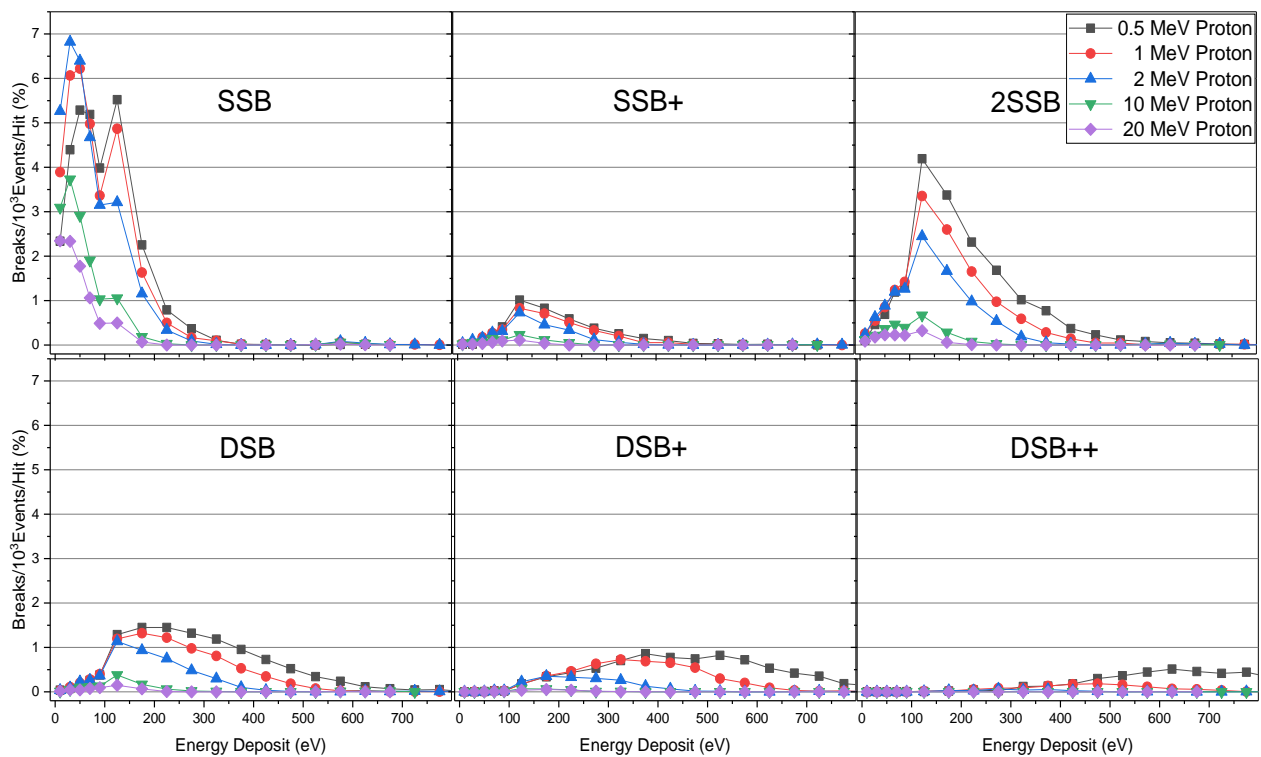

**Figure 5.** Number of breaks per hit for 1000 events, as function of the deposited energy in the DNA segment, for various proton energies.

**Table 4.** Total number of hits per gray in a DNA segment of 216 bp, obtained for the specified deposited energy intervals and primary proton energies. The results are reported in multiples of $10^{-5}$.

| Deposited Energy (eV) | 0.5 MeV Protons | 1 MeV Protons | 2 MeV Protons | 10 MeV Protons | 20 MeV Protons |
|---|---|---|---|---|---|
| **0 – 20** | 0.105 | 0.314 | 0.857 | 2.399 | 3.543 |
| **20 – 40** | 0.092 | 0.209 | 0.399 | 0.718 | 0.841 |
| **40 – 60** | 0.074 | 0.138 | 0.237 | 0.368 | 0.406 |
| **60 – 80** | 0.065 | 0.106 | 0.165 | 0.233 | 0.231 |
| **80 – 100** | 0.051 | 0.078 | 0.117 | 0.138 | 0.132 |
| **100 – 150** | 0.093 | 0.133 | 0.165 | 0.177 | 0.154 |
| **150 – 200** | 0.06 | 0.081 | 0.093 | 0.058 | 0.032 |
| **200 – 250** | 0.041 | 0.053 | 0.055 | 0.019 | 0.002 |
| **250 – 300** | 0.031 | 0.038 | 0.032 | 0.004 | |
| **300 – 350** | 0.024 | 0.03 | 0.017 | 0.002 | |
| **350 - 400** | 0.021 | 0.02 | 0.007 | | |
| **400 - 450** | 0.016 | 0.016 | 0.003 | | |
| **> 450** | 0.074 | 0.030 | 0.010 | | |

Figure 6 shows the average total number of SSB and DSB breaks per DNA segment, for different primary proton energies. The remarkable feature of the data is that for a given energy deposited in the DNA segment, there seems to be no significant dependence on the proton energy. This is the case especially for the results of the SSBs. Also to be noted is the saturation effect in the pattern of the SSBs at high LET values, where the probability of producing a break per unit energy deposition decreases with increasing energy deposition.

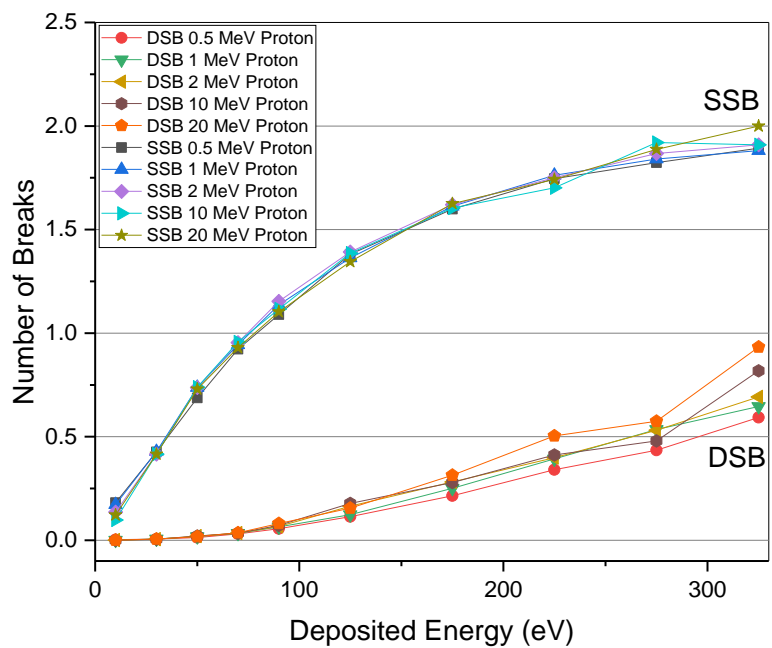

**Figure 6.** Average total number of produced single- and double-strand breaks per DNA segment as function of the deposited energy in a DNA segment of 216 nucleotides length. Here, the total number of single- and double-strand breaks are considered to be the sum of $SSB_{all}$ and $DSB_{all}$.

Table 5 compares the yield of the damage sites $(Gy.Gbp)^{-1}$, obtained from the total number of SSBs and DSBs, with the values obtained from the frequency of the deposited energies that are calculated as follows. Let n(E,y) be the number of breaks of particular type, say SSB or DSB, produced when energy

E is deposited in a DNA segment of length y per hit, and P(E,y) be the number of energy depositions of size E in length y per unit dose. Hence, n(E,y) would correspond to the values in figure 6, whereas P(E,y) correspond to the values in Table 4 (for example, for 10 eV deposited energy and 0.5 MeV protons, it is $0.105 \times 10^{-5}$). Following the procedure in Charlton *et al* [41], the yield can thus be calculated as $Yield = \sum n(E, y) \times P(E, y)/y$, in which the summation runs over the deposited-energy values and y = 216 is expressed in bp. The yield results of this calculation are shown in the 4th and 5th columns of Table 5. As can be seen, the results of these two columns reproduce the corresponding ones of the 2nd and 3rd columns reasonably well. The reason for such closeness of the outcomes is that in principle the two methods of calculation are the same; however, in employing the above mentioned formula of Charlton *et al* (1989), where we used the numbers in Table 4, we have chosen the mid values of each interval of deposited energy (say, 10 eV for the first interval) as E in the Yield formula. The yield values (Gy.cell)$^{-1}$ are also presented in this Table. In the derivation of the latter results, the average molecular weight of a chromosome was calculated by assuming 22 chromosomes per cell and a mean number of 245 Mbp per chromosome. The relative chromosome size distribution is taken from the cytometry techniques to be ranging from 610 Mbp to 40 Mbp with a number-averaged chromosome size of 245 Mbp [42] [12].

**Table 5.** Yield results as function of the proton energy. The results of the 2nd and 3rd columns are the same as in Table 2. The yield results of the 4th and 5th columns are extracted from the frequency of the deposited energies. The two columns on the right represent the yield values per gray for a cell with a number-averaged chromosome size of 245 Mbp, based on the results of the 4th and 5th columns.

| Energy (MeV) | $Y_{SSB}$ (Gy.Gbp)$^{-1}$ | $Y_{DSB}$ (Gy.Gbp)$^{-1}$ | $Y_{SSB}$ (Gy.Gbp)$^{-1}$ | $Y_{DSB}$ (Gy.Gbp)$^{-1}$ | $Y_{SSB}$ (Gy.cell)$^{-1}$ | $Y_{DSB}$ (Gy.cell)$^{-1}$ |
|---|---|---|---|---|---|---|
| **0.5** | 39.05 | 7.80 | 38.97 | 7.76 | 210.05 | 41.83 |
| **1** | 50.92 | 7.77 | 50.66 | 7.68 | 273.06 | 41.39 |
| **2** | 62.74 | 6.36 | 62.72 | 6.38 | 338.06 | 34.39 |
| **10** | 73.45 | 4.3 | 73.46 | 4.27 | 395.95 | 23.01 |
| **20** | 75.15 | 3.50 | 75.02 | 3.58 | 404.36 | 19.30 |

## Discussion

Introducing a primary proton source, various damage types in the DNA sample were calculated, based on the Geant4-DNA simulations for physical and chemical stages. For the physical stage, the threshold energy for recording a hit as a break was considered to be 17.5 eV. Same value has been used in the simulations by Meylan *et al* [18] where they simulated a fibroblast cell nucleus using Geant4-DNA, and Nikjoo *et al* [39] where they simulated B-DNAs using KURBUC. Using the PARTRAC code, Friedland *et al* [40] have investigated a threshold variation between 5 and 37.5 eV, implementing a linear acceptation probability (a linear increasing of the probability from zero, for a deposited energy less than 5 eV, to 1 when it exceeds 37.5 eV) for direct damage. In their simulations, they have implemented a basic chromatin fiber element including 30 nucleosomes and an ideal arrangement of chromatin fiber rods in rhombic loops forming a rosette-like structure of 0.5 Mbp genomic length. Whereas the interaction probability of the hydroxyl radicals is considered in Meylan *et al* [18] to be 0.4, we have adopted the value 0.13 which is the same as in Nikjoo *et al* [39] and Friedland *et al* [40]. Like in Nikjoo *et al* [39], we limited the chemical stage simulation time to 1 ns for the interaction of hydroxyl radicals with DNA – to be compared with 2.5 ns in Meylan *et al* [18]. In Geant4-DNA, the chemical stage is simulated in several time steps during which the movement of all the molecules is governed by their diffusion coefficients [36]. In our simulations, we did not specifically model the scavenging reactions that decrease the number of the existing hydroxyl radicals for damaging the DNA, whereas Friedland *et al* [40] has taken into account the scavenging of the chemical species at each time step due to random absorption of the radicals and as such considered an appreciably longer chemical stage simulation time of 10 ns.

It is already known that the cross sections used in the Geant4-DNA for the elastic, ionization, and excitation processes are less than the ones implemented in other codes [43] [44]. For instance, for electron energies higher than 100 eV, the difference between the excitation cross sections of CPA100 and Geant4-DNA amounts to about an order of magnitude [44]. Although the ionization cross sections of CPA100, compared to the other codes, agree better with the experimental ones, the ionization cross sections of Geant4-DNA and CPA100 are in reasonable agreement with each other, for electrons of more than 100 eV energy (produced numerously as secondary particles, having a primary proton source), where the ionization is the most important process [44] [17]. Furthermore, the maximum total excitation cross section in Geant4-DNA is smaller than in PARTRAC [45]. Table 1 shows that the chemical reaction rates of the hydroxyl radicals with other molecules and radicals (including hydroxyl) are less in Geant4-DNA as compared to the experimental values (see, for instance, the first row of Table 1), whereas the production rates of the hydroxyl radicals are more in Geant4-DNA as compared to experiment (see the third row of Table 1). Hence, in Geant4-DNA, the share of indirect damage is higher and more hydroxyl radicals are subject to reaction.

Despite the above differences between various simulations, the results of figure 3 with PARTRAC and KURBUC are comparable to our results and the ones represented in Meylan *et al* [18], the latter two using Geant4-DNA with different geometries. The reason for such agreement is partly due to using similar methodology in terms of the definition of SB with regards to the physical and chemical processes. Our simulations are in good agreement with the experimental data of Frankenberg *et al* [20] up to around 25 keV/μm as well as Belli *et al* [21] and Campa *et al* [23] over the whole range of LET shown in figure 3 (though there is only three data points available from the latter two experiments). This is especially true in regards with the data of Frankenberg *et al* [20] around a LET of 25 keV/μm, Belli *et al* [21] around 11 keV/μm, and Campa *et al* [23] around 28 keV/μm. Like the general trend of the experimental data, our DSB yield increases with the LET of the primary protons up to 7.8 (Gy.Gbp)[-1] for a LET of 39.7 keV/μm. Figure 3 also shows that the DSB yield increases with LET according to various simulations. Except for the DSB yield results of Nikjoo *et al* [39] which appear to show a linear increase with LET, our simulations as well as the simulations of Friedland *et al* [40] and Meylan *et al* [18] show a less pronounced increase of DSB yield above a certain value of LET (about 35 keV/μm for the latter two works and 25 keV/μm in our simulations). It is also interesting to note that among different simulations presented in figure 3, our results seem to show a better agreement of the trend and proximity to the experimental data points of Belli *et al* [21] and Campa *et al* [23].

**Conclusions**

In this article, using Geant4-DNA simulations and Python programming for data analysis, we presented our results for the frequency of simple and complex damage caused by primary protons in a B-DNA model. The simulation toolkit allowed us to calculate the energy depositions from physical interactions of protons, secondary particles, and account for the role of the hydroxyl radicals in producing strand breaks through chemical reactions. Simulations were performed introducing a point-source of protons at the center of a spherical liquid water medium [27] [13] [46] [39], isotropically thrown in full phase space with energies ranging from 0.5 MeV to 20 MeV. As such, the probability of simple and complex damages as well as SSB and DSB yields were calculated. The results were compared with other simulation works as well as the data obtained from experiments [20] [21] that used pulsed field gel electrophoresis. Overall, there is reasonable agreement between our results and the presented experimental data. Especially, our results as compared with other presented simulations, would give a better resemblance of the trend and value of the experimental data of Belli *et al* [21] and Campa *et al* [23]. The discrepancies observed between our DSB yield results and the corresponding ones in other simulation works can be assigned to the exploited DNA geometries, the type of chemical processes considered in the simulations, and different parameter adjustments and criteria for SB registration caused by physical or chemical processes.

## Acknowledgments

The authors gratefully acknowledge the Sheikh Bahaei National High Performance Computing Center (SBNHPCC) for providing computing facilities and time. The SBNHPCC is supported by the Isfahan University of Technology (IUT).